\documentclass{article}
\usepackage{fullpage}
\usepackage{times}
\usepackage{mathptm}
\usepackage{amsmath,amssymb,fancyhdr}
\usepackage{algorithm}                                  
\usepackage{algorithmic}
\usepackage{graphicx}
\makeatletter
\def\multilimits@{\bgroup
  \Let@
  \restore@math@cr
  \default@tag
 \baselineskip\fontdimen10 \scriptfont\tw@
 \advance\baselineskip\fontdimen12 \scriptfont\tw@
 \lineskip\thr@@\fontdimen8 \scriptfont\thr@@
 \lineskiplimit\lineskip
 \vbox\bgroup\ialign\bgroup\hfil$\m@th\scriptstyle{##}$\hfil\crcr}
\def\Sb{_\multilimits@}
\def\Sp{^\multilimits@}
\def\endSb{\crcr\egroup\egroup\egroup}

\makeatother

\usepackage{multirow}
\begin{document}

\begin{center}

{\LARGE\bf Lossless Layout Image Compression Algorithms
for Electron-Beam Direct-Write Lithography}

\end{center}

\begin{center}

{\large\bf Narendra Chaudhary, Yao Luo, Serap A.~Savari}

\end{center}

\noindent {\small\bf Dept.~of Electrical \& Computer Engineering,
Texas A\&M University, 3128 TAMU, College Station, TX 77843-3128}

\noindent {\bf Electronic mail: \underline{savari@ece.tamu.edu} }
\vspace*{.1in}

\begin{center}

{\large\bf Roger McCay}

\noindent {\small\bf GenISys, Inc.,
P.O.~Box 410956, San Francisco, CA 94141-0956}
 \end{center}
\vspace*{.1in}
\noindent Electron-beam direct-write (EBDW) lithography systems must 
in the future transmit terabits of information per second to be viable for
commercial semiconductor manufacturing.  Lossless layout image compression 
algorithms with high decoding throughputs and modest decoding resources
are tools to address the data transfer portion of the throughput problem.
The earlier lossless layout image compression algorithm Corner2 is designed
for binary layout images on raster-scanning systems.  We propose variations
of Corner2 collectively called Corner2-EPC and Paeth-EPC
which apply to electron-beam proximity corrected layout images
and offer interesting trade-offs between compression ratios and decoding
speeds.  Most of our algorithms achieve better overall compression
performance than PNG, Block~C4 and LineDiffEntropy while having low decoding 
times and resources.

\newpage

\subsection*{I. INTRODUCTION}

Electron-beam direct-write (EBDW) lithography is an attractive candidate for
next-generation lithography because electron beam lithography systems can 
attain very high resolution$^1$, and EBDW lithography systems do not need 
costly physical masks since the electron beam writer writes the mask pattern 
directly to the photoresist layer.
The widespread use of EBDW lithography in commercial semiconductor 
manufacturing has been hindered by its low throughput in the patterning of 
wafers.  In order to address this issue, future EBDW lithography systems 
must transmit terabits of information per second;  for example, 
Levinson$^2$ estimated that the patterning
of one 300-nm wafer per minute with 1-nm edge placement for 22-nm technology
requires the transmission of about 48.7 Tbits/s of data.
There have been endeavors to tackle the data transfer problem by using
massively parallel electron beams to write multiple pixels at a time$^{3-6}$.
However, multiple electron beam lithography systems are not currently ready
to replace optical lithography for high-volume manufacturing, and among the
problems to be addressed are datapath issues; i.e., there are questions
about how to provide the massive layout image data to the array of electron
beam writers. 

\begin{figure}
\centering
\includegraphics[width=8.382cm]{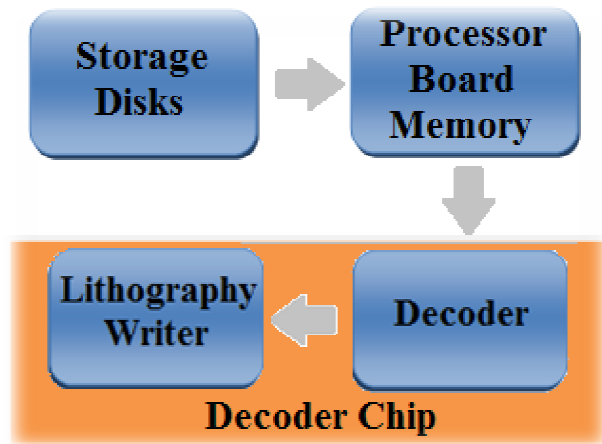}
\caption{(Color Online) The figure depicts data delivery for electron-beam
direct-write lithography systems. Reproduced with permission from 
J.~Vac.~Sci.~Technol., B~{\bf 32}, 06F502 (2014). Copyright 2014 American
Vacuum Society.}
\label{fig:dai}
\end{figure}

Dai and Zakhor$^{4,7}$ considered a datapath system which uses lossless data 
compression to approach this issue (see Figure~\ref{fig:dai}).  
Here, compressed layout images are cached in storage devices and sent to the 
processor memory board.  To satisfy the throughput requirements,
the decoder embedded within the array of electron beam writers
must be able to rapidly recover the original layout images from the
compressed files. 

The Block~C4 algorithm$^4$ (BC4) is intended for a raster 
scanning system in which each pixel is quantized to one of 32 levels. 
This lossless layout image compression algorithm is based on a combination
of context prediction and finding repeated regions within a layout image.
The Block~C4 algorithm inspired other lossless layout image compression
algorithms including those by Cramer et al.$^8$ and Carroll et al.$^9$
for the reflective electron beam lithography (REBL)$^5$ system. 

There have been other lossless layout image compression algorithms
for the datapath system of Dai and Zakhor which utilize different techniques
than the Block~C4 algorithm and related algorithms.  
The most recent of these$^{10}$ proposes a hybrid of standard data compression
algorithms that are widely applied outside of lithography.
In this paper we will focus on a collection of gray-level layout image
compression algorithms that are based on or have similarity to the lossless
compression scheme Corner2$^{11,12}$ for binary layout images on 
raster-scanning systems.  The Corner2 algorithm is based on a transform which 
exploits the fact that most polygons in layout images have right angle 
corners. The Corner2 decoder can reproduce layout images with only a small 
cache, and Corner2 and its precursor Corner$^{13}$ were found to have better 
compression performance than the Block~C4 algorithm.  The earlier gray-level
layout image compression algorithms that are the most similar to Corner2
are CornerGray$^{14}$ and LineDiffEntropy$^{15}$, and we will discuss them
in Section~II.B.

Krecinic et al.$^{16}$ independently proposed a vertex-based circuit layout
image representation format, which can be viewed as a variant of the Corner2
algorithm.  However, Corner2 also incorporated ways to take advantage of
regularity within circuits as well as advanced entropy encoding techniques
to further compress the representation;  the algorithm of Krecinic et al.~did 
not include similar features.

In this paper, we propose a collection of schemes called Corner2-EPC 
and Paeth-EPC which
extend the image transformation technique of Corner2 to handle pixels which
can be quantized to an arbitrary number of gray levels; we used 32.
The layout images used in our experiments were processed with the electron 
beam proximity effect correction algorithm of GenISys, Inc.,
{\small BEAMER}\_v4.6.2\_x64.  We briefly comment on the range of values
used by the proximity effect correction algorithm and the control of edge
placement.  High-speed electron-beam direct-write lithography  is intended 
for the high-volume manufacturing of semiconductors and therefore to standard 
Si and SiO$_2$ substrate materials.  The dynamic dose range of these materials 
is generally 0.75 to 1.45 for a range of pattern densities.  This range is 
within a dose factor of 2.  If the 32nd dose is set to the dose factor of 
1.45, then the 0.75 dose would fall in about the middle of the dose step 
range leaving the bottom half of the 32 dose steps available for edge 
movements of the lowest dosed features, which would appear to be the worst 
case.  If a 22 nm writing pitch was being used for exposures that would 
potentially give the control of about 1.4 nm edge movements, but this is an
incomplete discussion.  The doses of less than 1.00 are for the interiors of 
large shapes, which have no bearing on the critical dimensions or edge
placements.  The interior of shapes do not need the granularity of adjustment,
they just need enough dose to clear.  Therefore, all of the doses below 1.00 
can be used for edge movements, which gives some 24 of the 32 dose steps, for 
a writing pitch of 22nm that results in sub-1 nm edge control.

Most of the compression schemes we discuss attain better
overall compression performance than the standard image compression algorithm
PNG$^{17}$ as well as  Block C4 and LineDiffEntropy while having modest
decoding time and complexity.  Although we are presenting 
results for a software implementation, Corner2-EPC has been designed so that 
it is amenable to hardware implementation.  Paeth-EPC can also be implemented
in hardware, but it has slower encoding and decoding times so that it can
attain better compression ratios.

The remainder of the paper is organized as follows.  In Section~II we present
more information about some of the earlier work on layout image compression
algorithms.  In Section~III we introduce the Corner2-EPC schemes and their
variants, the Paeth-EPC schemes.
In Section~IV we provide experimental results for two circuits.

\subsection*{II. ON LOSSLESS LAYOUT IMAGE COMPRESSION}
\subsubsection*{{\em A. Some Algorithms Related to the Lempel-Ziv Codes}}
The earliest lossless data compression schemes to be devised were for
one-dimensional data.$^{18}$
Entropy encoding and entropy decoding respectively refer to the encoding 
and decoding/decompression of one-dimensional data, and
image compression/decompression schemes often incorporate
entropy encoding and decoding algorithms$^{17}$.  

The Lempel-Ziv codes$^{17-20}$ are among the most widely used lossless data
compression schemes. 
The original Lempel-Ziv codes$^{19,20}$ date back to the 1970's, and there 
are numerous variants of these codes$^{18}$ which are the foundation
for many practical compression schemes because they have simple 
implementations, their decoding is comparatively fast, and they often need 
comparatively little memory.$^{17}$  
The Lempel-Ziv schemes and their variations encode
variable-length segments of data by pointing to the preceding appearance of
each segment; the segments are kept in a dictionary which is being continually
updated according to the specific algorithm.  

One of the ways to assess the efficiency of a lossless layout image
compression algorithm is to compare its performance to that of a general
purpose image compression algorithm.  The one we will use for this purpose
is Portable Network Graphics (PNG)$^{17}$.  The compression algorithm
utilized by this Internet standard is based on {\it deflate}$^{21}$,
which is one combination of the Lempel-Ziv code LZ'77$^{19}$ with the
Huffman code$^{22}$ or an approximation to it; the Huffman code is another 
famous entropy code.

The Block C4$^4$ algorithm was one of the lossless layout image compression
algorithms proposed by Dai and Zakhor.  The algorithm is based on a 
two-dimensional version of a Lempel-Ziv code and also incorporates context 
prediction.  The algorithms discussed in References~8 and 23-24 are
refinements to this compression scheme.
Carroll et al.$^9$ developed a lossless layout image compression algorithm
which they implemented in the digital pattern generator of the REBL system.
Although the exact details of the scheme were not provided 
it appears to be computationally simpler than the Block C4 algorithm
as the authors were concerned with issues like
space constraints, operating speed, and power consumption.  
The authors of Ref.~9 segment layout images into `triLines,' which are groups
of three rows; each ``superpixel'' of a triLine consists of
a column of three pixels from the three rows which have been aggregated.
These triLines are encoded by a hybrid of Lempel-Ziv-like compression 
with run-length encoding$^{25}$ and repeat and edit features.  
This approach accounts for
some of the data dependencies between successive rows although
it generally cannot handle many dependencies between 
pixels.  This scheme more closely resembles a one-dimensional version of a
Lempel-Ziv code than a two-dimensional variant of a Lempel-Ziv code.  
Carroll et al.~reported that the compression ratio and throughput of
the decoder needed were not consistently achieved although they had success
with ``specially contrived patterns... such as line-space patterns and via
patterns whose pitch is compatible with the raster pixel-pitch.'' 

While the Lempel-Ziv codes are popular for rapid implementations in designs 
where decoding speed and memory are important, this popularity is due in part
to the ability of the Lempel-Ziv codes to attain some compression on a large 
collection of data types.  Hence, it is interesting to consider 
compression schemes which may better match certain features of the lossless
layout image compression problem.

\subsubsection*{{\em B. Some Algorithms Related to the Corner2 Algorithm}}

The lossless compression scheme Corner2$^{11,12}$ 
for binary layout images on  raster-scanning systems
was motivated by the GDSII$^{26}$ format, which is one of the standard formats
for storing circuit layouts.  The GDSII format represents circuit features
such as polygons and lines by their corner points.
Circuit data which is described in GDSII format is far more compact than
the uncompressed image of a circuit layer. 
However, the GDSII format is not well-suited for EBDW applications because
it takes hours on a complex computer system with large memory to convert
a GDSII representation into the layout images needed for the lithography
process.  The Corner2 algorithm is designed to take advantage of the idea
of corner representation while avoiding the complex processes used to convert
a GDSII file into layout images. 

We can summarize the operation of the Corner2 algorithm as follows.
The algorithm attempts to account for some of the regularity within the circuit
by creating a dictionary of frequently occurring patterns and modifying the
layout image to replace these patterns with pointers to them.
The next step is a corner transformation process on the modified image
to create a third image;  we will consider extensions of this transformation
in Section~III.  This transformation is based on the observation that a
pixel in a layout image often has the same value as the previous pixel in
the horizontal direction and/or the previous pixel in the vertical direction.
This third image generally contains many pixels with value
zero.  This third image is processed with a combination of run length 
encoding$^{25}$ and end-of-block coding to obtain a one-dimensional data
stream; we will discuss run length encoding and end-of-block coding in
Section~III.  The final step in the Corner2 encoding of a layout image is to
apply arithmetic coding$^{17, 18, 27-30}$ to the one-dimensional data stream;
this step is one of the options we will consider in Section~III.
The Corner2 algorithm has a simple decoding process.  

Most of the earlier extensions of the Corner2 algorithm were intended
for binary layout images.  Reference~$31$ improves upon the frequent pattern 
replacement portion of the Corner 2 algorithm.
References~32 and 33 discuss Corner2-MEB, which modifies the Corner2 
algorithm to account for the lattice
formation of the positions and the zigzag writing strategies of the electron
beam writers in the MAPPER$^6$ system.

Corner2 greatly outperformed the Block C4 algorithm
in compression
ratio, encoding/decoding times and memory usage, but it did
not account for the properties of the objects in a layout image that has
been processed by software for electron-beam proximity effect corrections.
While the input to the Block C4 algorithm is a gray layout image the discussion
about Block C4 only considered edge placement and not proximity effect
corrections.   The CornerGray$^{14}$ algorithm was the first attempt
to extend the Corner2 algorithm to gray-level images.  It omits the frequent
pattern replacement step of the Corner2 algorithm and assumes the gray-level
layout images have objects with pixel intensities that are fully filled 
inside of polygon outlines, empty outside of polygon outlines, and taking
on intermediate values along polygon outlines which have a uniform depth
of one pixel.  The transformation step of the CornerGray algorithm 
produces a ``corner stream'' which identifies the horizontal/vertical
transition points and an ``intensity stream'' which represents the 
corner/edge intensities.  These streams are processed by separate entropy
encoders.  The encoding of the ``corner stream'' closely resembles most of
the steps of the Corner2 transformation.  The encoding procedure for the
``intensity stream'' uses information about the run lengths of a single 
intensity
value along a horizontal edge or along a vertical edge in determining
the description of the intensity stream.  The output of that process is then
compressed by a combination of the LZ '77 algorithm and the Huffman code.
The experimental results reported on a 500~nm memory core with repeated
memory cell structure were a slight improvement in overall compression ratio
and significant improvements in encoding and in decoding times over the
Block~C4 algorithm.

The LineDiffEntropy$^{15}$ algorithm is a gray-level layout image
compression algorithm which is also designed to take advantage of the
fact that a pixel in a layout image often has the same value as the previous
pixel in the horizontal direction and/or the previous pixel in the 
vertical direction.  The authors of Reference~15 provide an explicit
description of the compression and decompression schemes they have proposed.
As with the Corner2-MEB algorithm, the LineDiffEntropy algorithm segments the
layout image into blocks which correspond to the writing region of a single
electron beam writer.
Like the Corner2 algorithm and its variants, the LineDiffEntropy encoder and
decoder operate in a row-by-row fashion.  However, the LineDiffEntropy 
algorithm does not use a transformation like the Corner2 algorithm or the
variations we will propose in Section~III.  Instead, the LineDiffEntropy 
algorithm uses string matching to determine regions where a row of pixels
is identical to the previous row and for other regions it describes pairs
of intensity values and run lengths of those intensity values.  Like the
Corner2 algorithm and its variants, the LineDiffEntropy algorithm uses a
symbol that is essentially identical to the End-of-Block symbol of 
the Corner2 algorithm, but it differs from the Corner2 algorithm in that
it does not use run length encoding on the End-of-Block symbols.
There is a compaction step to make the resulting description of sequences
of rows more succinct, and there are 1056 different possible symbols 
that can be used by the encoder up to this point.  Each of these 1056
symbols is represented by a binary codeword of a prefix condition code.
The authors report improvements in compression ratios and significant
improvements in encoding and decoding times over the Block~C4 algorithm. 

\subsection*{III. MODELLING AND ALGORITHMS}
We can outline the operation of the Corner2-EPC compression algorithms as 
follows.  First, each layout image goes through a transformation.  Second, 
we may elect to use an additional entropy coding step to attain further
compression.  In the next two subsections we will offer more details
about these steps.  In the third subsection we will describe how to invert
the image created during the transformation to reproduce the original layout
image.  In the fourth subsection we will describe the Paeth-EPC variation, 
which alters the first step of the Corner2-EPC transformation to attain 
improved compression ratios at
the expense of increases in encoding and decoding times.

\subsubsection*{{\em A. The Corner2-EPC Transformation}}
\begin{figure}
\centering
\includegraphics[width=16cm]{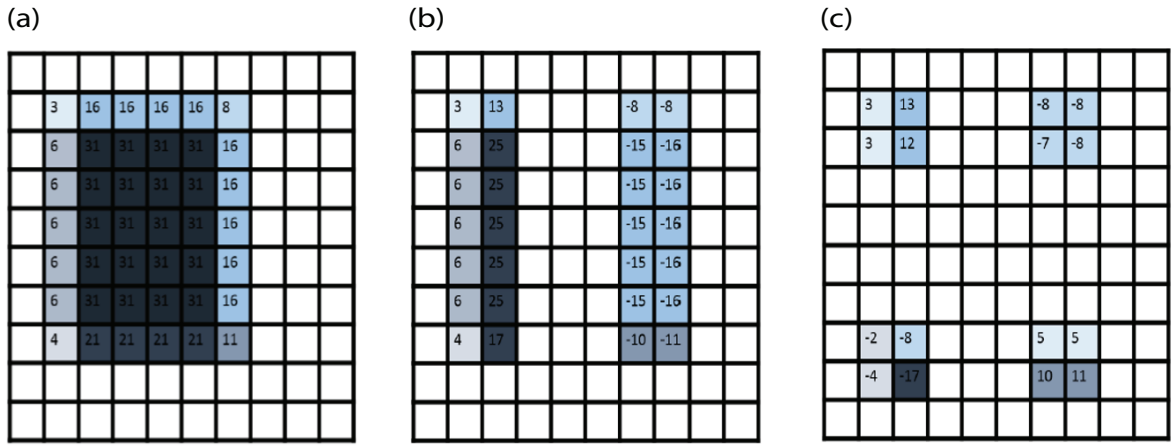}
\caption{(Color Online) The figure illustrates the two-step transformation 
of an image. (a) The original image has pixel values in 
the range $\{0, \dots , 31\}$. (b) After the horizontal coding step, the 
intermediate image has pixel values in the range $\{-31, \dots , 31\}$. 
(c) After the vertical coding step, the transformed image has pixel values 
in the range $\{-62, ..., 62\}$.}
\label{fig:trans}
\end{figure}

A pixel in a proximity corrected layout image frequently has the same value
as the previous pixel in the horizontal direction and/or the previous
pixel in the vertical direction.  Figure~\ref{fig:trans} illustrates a 
two-step 
transformation in which the intermediate image is obtained by replacing
each pixel in a row after the first one with the difference between the
pixel value the preceding pixel value from the original image.
The final image is obtained by replacing each pixel in a column after the 
first one with the difference between the pixel value the preceding pixel 
value from the intermediate image.
Although it is conceptually simple to describe this transformation as a
two-step process it is more effective to implement it in one step, which
we summarize in Algorithm~1.  In the algorithm, $x$ denotes the column index 
$[1, \cdots, C]$ of the image and $y$ denotes the row index 
of the image $[1, \cdots, R]$.  Observe that if the pixels in the original
layout image take on values in the set $\{0, \ 1, \ \dots , \ 31\}$, then
the pixels in the Corner2-EPC transformed image take on values in the set 
$\{-62, \ -61, \ \dots , \ 61 , \ 62 \}$.  

\begin{algorithm}[!htb]
\caption{Corner2-EPC Image Transformation : One-Step Algorithm}
\label{alg:transform1}
\begin{algorithmic}[1]
\renewcommand{\algorithmicrequire} {\textbf{Input:}}
\REQUIRE Layer image $\texttt{IN} \in \{0, 1, \dots , 31\}^{C \cdot R}$
\renewcommand{\algorithmicrequire} {\textbf{Output:}}
\REQUIRE Corner image $\texttt{OUT} \in \{-62, -61, \dots , 61, 62 \}^{C \cdot R}$
\STATE Initialize $\texttt{OUT}(1,1)= \texttt{IN}(1,1).$
\FOR {$x=2$ \TO $C$} 
        \STATE $\texttt{OUT}(x,1)= \texttt{IN}(x,1) - \texttt{IN}(x-1,1).$
\ENDFOR
\FOR {$y=2$ \TO $R$}
        \STATE $\texttt{OUT}(1,y)= \texttt{IN}(1,y) - \texttt{IN}(1,y-1).$
        \FOR {$x=2$ \TO $C$}
        \STATE $\texttt{OUT}(x,y)= \texttt{IN}(x,y) + \texttt{IN}(x-1,y-1) - \texttt{IN}(x-1,y) - \texttt{IN}(x,y-1).$
        \ENDFOR
\ENDFOR
\end{algorithmic}
\end{algorithm}

As Figure~\ref{fig:trans} suggests, the transformed image tends to have many
zeroes, and it is therefore effective to represent it with some form of 
run length encoding$^{25}$.  The idea in the simplest version of run-length
encoding is to describe nonzero values as they are but to count the repeated
zero values between successive nonzero values and to describe each ``run''
of zero values with an encoding of its run length.  Run length encodings
are widely used$^{17}$, and there are many variable-length representations
of the integers$^{25,34}$. 
Run length codes such as the Golomb-Rice code and the
Exponential-Golomb code are components of video standards.$^{35}$
Moussalli et al.$^{36}$ offer an FPGA implementation of a Golomb-Rice decoder
with 7.8~gigabits per second throughput using 10\% of the available resources
of a mid- to low-size FPGA.
In 2014, a typical Golomb-Rice decoder implemented as part of a video decoder
had a throughput of 10 gigabits per second on a 500 Mhz/16 nm process and
7.6 gigabits per second on a 380 Mhz/28 nm process$^{37}$.

The form of run length encoding that we consider here is taken from the
Corner2 algorithm and its variants.  Our data initially takes on values in
the range $\{-62, \ -61, \ \dots , \ 61, \ 62\}$.  The first step is to
segment the data into blocks of a predetermined length $L$.
In the example below, let $0^i$ denote a sequence of $i$ zeroes.
Suppose $L=7$ and our initial data sequence is segmented as
\begin{displaymath}
20 \ 0^6 \hspace*{.2in} 0^4 \ 16 \ 0^2 \hspace*{.2in} 0^7
\hspace*{.2in} -15 \ 0^6 \hspace*{.2in} 0^7.
\end{displaymath}
In the next step of converting this stream, we introduce a temporary 
``End-of-Block''
symbol $X$ which indicates that the remainder of the block is a sequence of
zeroes.  Our intermediate representation of the preceding data sequence
becomes
\begin{displaymath}
20 \ X \hspace*{.2in} 0^4 \ 16 \ X \hspace*{.2in} X
\hspace*{.2in} -15 \ X \hspace*{.2in} X.
\end{displaymath}
For our final representation of the data we introduce $M+N$ symbols
and remove the earlier symbols $0$ and $X$.  $M$ of the symbols denote
the base-$M$ symbols ``$0_M$'', ``$1_M$'', $\cdots$ , ``$(M-1)_M$'',
and for each $1 \leq i < L$ we replace every occurrence of $0^i$ with the
base $M$ representation of $i$ using these $M$ symbols.
A typical intermediate data sequence has (sometimes long) runs of ``X''s.
Our other $N$ symbols denote
the base-$N$ symbols ``$0_N$'', ``$1_N$'', $\cdots$ , ``$(N-1)_N$'',
and we replace each run of ``X''s with the base $N$ representation of its
run length using these $N$ characters.  In a hardware implementation, it may
be necessary to restrict the maximum length of runs of ``X''s that can be
decoded at once.

Observe that there are $124+M+N$ possible symbols used to represent the
transformed layout image.  For ease of implementation it is preferable to
select $M$ and $N$ to be powers of two. As we will explain in the next 
subsection, it is
convenient to choose $124+M+N \leq 255$ to take advantage of compression
algorithms that are available online.  However, for one 
Corner2-EPC scheme we examine we assign an 8-bit string to each of the
$124+M+N$ possible symbols, and we use no further encoding.

Although we have separately discussed the image transformation algorithm
and the run length encoding algorithm, these are implemented together.
If an additional entropy encoding step is used, then we are handling it
separately to use open source implementations.

References~32 and 33 discuss
how to adapt the Corner2 transformation to handle a zigzag writing order.
Algorithm~\ref{alg:transform1} and its inverse transformation and the
variation later given for Paeth-EPC can be
similarly adapted for a zigzag writing order, but we omit the details here.

\subsubsection*{{\em B. Additional Entropy Encoding}}
As we will discuss in Section~IV, we attain better overall compression
ratios than the PNG image compression standard without additional entropy
encoding.  However, the data stream produced by the algorithm of the previous
subsection can be compressed further, and it is interesting to consider ways
to do so which are amenable to hardware implementations.  As with the
Corner2 algorithm and its variants, we consider arithmetic coding.
Arithmetic coding is widely implemented in video coders and 
decoders$^{28-30,35}$.  We use an open source implementation
which permits a maximum input symbol alphabet size of 256.

There have been hardware implementations of the Lempel-Ziv codes targeting
high throughput which are motivated by the needs of data centers and
communication networks$^{38}$.  IBM researchers$^{39}$ have sustained
throughputs of 3 gigabytes per second$^{40}$ for gzip (which combines
LZ '77 and the Huffman code or an approximation to it) on the Canterbury
corpus and the Large corpus$^{17}$.  The AHA3642 integrated circuit
offers a 20 gigabit per second compression/decompression throughput and is
described as ``low cost.''  Because deflate is the basis for gzip,
we use the standard {\em zlib} implementation$^{21}$ of deflate
which permits a maximum input symbol alphabet size of 256.

\subsubsection*{{\em C. Inversion of the Corner2-EPC Transformation}}
Just as the Corner2-EPC Transformation involves a transformation of a
layout image and some form of run length encoding, the inversion of this
process requires the corresponding run length decoding and the inverse
transformation of an image.  Algorithm~\ref{alg:inverse1} summarizes
the inverse image transformation algorithm.

\begin{algorithm}[!htb]
\caption{Corner2-EPC Inverse Image Transformation}
\label{alg:inverse1}
\begin{algorithmic}[1]
\renewcommand{\algorithmicrequire} {\textbf{Input:}}
\REQUIRE Corner image $\texttt{IN} \in \{-62, -61, \dots , 61, 62 \}^{C \cdot R}$
\renewcommand{\algorithmicrequire} {\textbf{Output:}}
\REQUIRE Layer image $\texttt{OUT} \in \{0, 1, \dots , 31\}^{C \cdot R}$
\STATE Initialize $\texttt{OUT}(1,1)= \texttt{IN}(1,1).$
\FOR {$x=2$ \TO $C$} 
        \STATE $\texttt{OUT}(x,1)= \texttt{IN}(x,1) + \texttt{OUT}(x-1,1).$
\ENDFOR
\FOR {$y=2$ \TO $R$}
        \STATE $\texttt{OUT}(1,y)= \texttt{IN}(1,y) + \texttt{OUT}(1,y-1).$
        \FOR {$x=2$ \TO $C$}
        \STATE $\texttt{OUT}(x,y)= \texttt{IN}(x,y) - \texttt{OUT}(x-1,y-1) + \texttt{OUT}(x-1,y) + \texttt{OUT}(x,y-1).$
        \ENDFOR
\ENDFOR
\end{algorithmic}
\end{algorithm}

Just as we implement a combination of the image transformation algorithm
and the run length encoding algorithm, we implement the run length decoding
algorithm together with the inverse image transformation because it is
impractical to store entire images.  As Algorithm~\ref{alg:inverse1} indicates,
a layer image can be decoded on a row-by-row basis.  If the decoder
is currently working on row $y$, it uses the output of the run length
encoding step to recover row $y$ of the transformed image as well as the
previous row and current row of the original layer image.

In terms of decoding an additional entropy encoder, we can decode that
last step separately, and this is the approach we use for deflate.
For arithmetic coding we adopt a ``memory save'' mode in which arithmetic
decoding  is combined with all of the other decoding operations.

\subsubsection*{{\em D.~The Paeth-EPC Transformation}}
The first step of the Corner2-EPC transformation inputs an image with
pixels taking on values $\{0, \ 1, \ \dots , \ 31\}$
and outputs a sparse image with pixels taking on values
$\{-62, \ -61, \ \dots , \ 62\}$.
We can alternatively use a computationally more demanding scheme motivated
by the Paeth$^{41}$ filter which produces another sparse output image but
has pixels taking on values $\{0, \ 1, \ \dots , \ 31\}$.
In addition to the input image $\texttt{IN}(x,y)$ and output image 
$\texttt{OUT}(x,y)$, the first step of the Paeth-EPC transformation
must also maintain a ``prediction'' image $\texttt{PRED}(x,y)$ which
has pixels taking on values $\{0, \ 1, \ \dots , \ 31\}$ and is defined as
follows.
\begin{eqnarray*}
\texttt{PRED}(1,1) & = & 0 \\
\texttt{PRED}(x,1) & = & \texttt{IN}(x-1,1) , \; \mbox{if $x \geq 2$} \\
\texttt{PRED}(1,y) & = & \texttt{IN}(1,y-1) , \; \mbox{if $y \geq 2$} \\
\texttt{PRED}(x,y) & = & \mbox{the closest among} \;  
\texttt{IN}(x-1,y), \texttt{IN}(x,y-1), \; \mbox{and} \; \texttt{IN}(x-1,y-1) 
\\ & & \mbox{to} \; 
\texttt{IN}(x-1,y) + \texttt{IN}(x,y-1) - \texttt{IN}(x-1,y-1) 
\; \mbox{if} \; x \geq 2, \ y \geq 2.
\end{eqnarray*}
Although the prediction image is explicitly provided by the Paeth filter,
the way to use it to produce an output image is not.  For computational
simplicity we choose
\begin{displaymath}
\texttt{OUT}(x,y) \equiv (\texttt{IN}(x,y) - \texttt{PRED}(x,y)) \bmod 32 \;
\mbox{for all $x,y$}.
\end{displaymath}

At the decoder, suppose the input pixel is $\texttt{IN}_d (x,y)$
and the decoder wishes to recover the original pixel  $\texttt{OUT}_d (x,y)$.
Then the decoder will use the earlier decoded pixels $\texttt{OUT}_d (x-1,y),
\texttt{OUT}_d (x,y-1), \; \mbox{and} \; \texttt{OUT}_d (x-1,y-1)$ to
calculate the value of $\texttt{PRED}(x,y)$ and will subsequently compute
\begin{displaymath}
\texttt{OUT}_d (x,y) \equiv (\texttt{IN}_d (x,y) + \texttt{PRED}(x,y)) 
\bmod 32 \; \mbox{for all $x,y$}.
\end{displaymath}

We use the same approach to run length encoding as we do for the Corner2-EPC
transformation.  As we will see in the next section, this approach leads to
better compression ratios at the expense of encoding and decoding times.

\subsection*{IV. RESULTS AND DISCUSSION}
There are two circuits for which we provide experimental results.
The experiments were performed on Intel i7-2600 CPU 
processors at 3.40 GHz with 8 GB of RAM using a WD Elements 1~TB portable
external drive, a Windows7 Enterprise operating
system and the electron beam proximity effect correction algorithm of 
GenISys, Inc., {\small BEAMER}\_v4.6.2\_x64.  
The implementations of the algorithms we propose and LineDiffEntropy (LDE)
are written in C/C++.  The implementation we received for the Block~C4
algorithm (BC4) is in C$\sharp$.

The output of the {\small BEAMER} software is in PNG format, which
initially represents the value of each pixel with eight bits instead of five.
Therefore the input to all of the algorithms is in initially in 
PNG format, but this input is subsequently converted to pixels.
We define the compression ratio of a layer as
 $$\frac{\texttt{Original Image File Size}}{\texttt{Compressed File Size}}.$$ 
The file sizes contain any overheads needed by the decoder and are all
measured in bytes; for example,
we used four bytes each to represent the width of the image, the height of the
image and the length of the data stream produced at the end of the
run length encoding step.
The last row of Tables~\ref{table:compression-denny} 
and \ref{table:compression-chipframe} 
is not the average of the preceding rows in the respective tables, but instead
$$\frac{\texttt{Total Original File Size}
}{\texttt{Total Compressed File Size}}.$$
With the exception of the LineDiffEntropy results, our encoding time 
measurements include the write time of the compressed file to the portable drive,
and our decoding time measurements include the read time of the compressed file
from 
the portable drive.  We make an exception with the LineDiffEntropy results
because we are not using the original code for this algorithm.

The first circuit we consider is a 25-layer image compression block based on 
the FREEPDK45 45nm library with a minimum element of 60 nm.  
We use a pixel size of 30~nm$\times$30~nm, and each layout image consists of
30403$\times$30324 pixels.
For both circuits, we experience a memory shortage for the encoding 
process when we attempt to run BC4 on an entire layout
image and hence have to segment the image into the largest components for 
which BC4 could be applied. 
In the case  of the image compression block
we split each layout image into four segments which are 
approximately quadrants of the image.
The experiments for all other algorithms were on full layout images.  

The second circuit we study is a 18-layer binary frequency shift keying
(BFSK) transmitter targeting 250~nm lithography technology.
We use a pixel size of 40~nm$\times$40~nm, and each layout image consists of
79050$\times$79050 pixels.
For the BC4 experiments we split each layout image into twenty-eight segments
where each segment consists of about one quarter of the rows and
one seventh of the columns.

We briefly comment on the proximity effect correction algorithm we used
and the dose assignments for primitives. 
In the case of the layout images used for the experiments most of the shapes 
were of such a small size that a single calculated dose was applied to the 
shape and did not require fracturing into smaller primitives for dose 
adjustment.  If a shape is large enough and is influenced by other nearby 
shapes the GenISys proximity effect correction algorithm {\small BEAMER}
will physically fracture it into smaller primitive shapes to achieve the 
proper doses for maintaining the critical dimension and the edge placement 
accuracy of the whole original shape.  The proximity effect correction
algorithm is not applied on a pixel-by-pixel basis.  When the moving of 
edges was required due to differences between the design grid and the writing 
grid this was performed after the proximity effect correction algorithm was 
applied, and this could generate an additional one to eight primitives which 
could be as small as one pixel.

The earlier image compression algorithms we consider in our experiments are
the Portable Network Graphics standard (PNG),
the Block C4 algorithm (BC4), and
the LineDiffEntropy algorithm (LDE).
For the four algorithms of ours for which we report results, we 
use the parameters $M=N=64$ and set $L$ to be the number of pixels in a row of
the layout image.
We report results for
the Corner2-EPC algorithm without additional entropy encoding
(Corner2-EPC (plain)),
Corner2-EPC algorithm followed by arithmetic coding 
(Corner2-EPC (AC)),
the Corner2-EPC algorithm followed by deflate (Corner2-EPC (deflate)), and
the Paeth-EPC algorithm followed by deflate (P-EPC (deflate)).

Tables I and II provide the compression ratios and a summary of 
encoding and decoding time statistics for the image compression block,
and Tables~III and IV
provide the corresponding results for the BFSK transmitter circuit.
We include the maximum of the decoding times among layers because this worst 
case may be important in a production environment.
In all cases we provide an approximation of the PNG encoding and decoding 
times as well as an underestimate of
the LineDiffEntropy encoding and decoding times.  The PNG ``encoding'' time 
of an image is the time used by the libpng algorithm to write the 
compressed file to disk one row at a time, and the ``decoding'' time is the 
time used by the libpng algorithm to read the compressed file from the disk
5000 rows at a time.  The LineDiffEntropy
encoding and decoding times do not include the times to respectively write
and read from the disk.

In terms of compression ratios, the algorithms we propose using deflate
outperform all other algorithms on every layer for both circuits.
The original Lempel-Ziv codes are known$^{20,42}$ to 
asymptotically attain maximum data
compression rates for certain types of one-dimensional data, but there are
no comparable results for their performance on two-dimensional data.
Although we do not use a frequent pattern replacement step in the proposed
algorithms that resembles the one used in Corner2, the deflate algorithm
effectively has a similar role.  We also observe that most of the compression
from our proposed algorithms comes from the first transformation step.

In terms of encoding times, we provide an approximation for the PNG algorithm
and an underestimate for the LineDiffEntropy algorithm.  
The LineDiffEntropy algorithm has some
additional advantage over the proposed algorithms because it does not use
an image transformation technique and it has one run-length encoding part
instead of two.  However, the algorithms we provide have reasonable
encoding times.

Decoding times are important for the electron-beam direct-write application.
Here the Corner2-EPC algorithms have the best performance in the worst-case
decoding time of a layer, and for the BFSK transmitter the deflate version
is also the best for total decoding time among all layers.

As we suggested earlier, the Paeth-EPC algorithms improve the compression
ratios of the Corner2-EPC algorithms at the expense of encoding and decoding
times.  The individual compression gains of both Paeth and deflate are 
greater in the BFSK transmitter circuit than in the image compression block;
we speculate that this is partly because there is more alignment of patterns 
within the BFSK transmitter circuit than within the image compression block.

\subsection*{V. SUMMARY AND CONCLUSIONS}
We have presented a group of layout image compression algorithms that offer
high compression ratios and show trade-offs between compression ratios
and decoding times.  The Block~C4 algorithm offers relatively uniform
compression ratios among layers.  The LineDiffEntropy algorithm offers
low decoding times for sparse layers.  The Corner2-EPC algorithm with no
additional entropy encoding appears to have the simplest decoding in terms
of memory requirements.  The Corner2-EPC algorithm with arithmetic coding
offers a compromise between simplicity of decoding and compression ratios.
The Corner2-EPC algorithm with deflate offers the best compression ratios
without compromising decompression time, but it may have more memory 
requirements than the previous algorithms.  The Paeth-EPC algorithms offer
the best compression ratios at the expense of decoding time and memory.

For future research, Lin$^{43}$ discussed the need
to add redundancy to layout image data being written to the REBL system
to reduce sensitivity to contamination on the digital pattern generator;
it would be desirable to understand how this issue affects the data delivery
problem for the REBL system

\subsection*{{\em Acknowledgments}}
This work was supported in part by NSF Grant No.~ECCS-1201994
and was made possible through an Agreement of
Cooperation between GenISys, Inc. and Texas A\&M Engineering Experiment
Station at College Station.  The authors also thank S.~Khatri, D.~Lie, and
S.~Mukhopadhyay for their circuit data and V.~Dai for providing an
implementation of the Block~C4 algorithm.  The authors are grateful to
J.~Yang for providing them with an implementation of the Corner2 algorithm
and for other helpful correspondences.

\noindent $^1$T.~R.~Groves, {\em Nanolithography: The Art of Fabricating 
Nanoelectronic and Nanophotonic Devices and} \\
\indent {\em Systems}, edited by
M.~Feldman, (Woodhead Publishing, Philadelphia, 2014), pp.~80-115.

\noindent $^2$H.~J.~Levinson, {\em Principles of Lithography}, 3rd ed.
(SPIE, Bellingham, 2010), p.~466.

\noindent $^3$N.~Chokshi, D.~S.~Pickard, M.~McCord, R.~F.~W.~Pease, Y.~Shroff,
Y.~Chen, W.~G.~Oldham, and D.~Markle, \\
\indent J. Vac. Sci. Technol., B {\bf 17}, 3047 (1999).

\noindent $^4$V.~Dai, ``Data compression for maskless lithography systems: Architecture, algorithms, and \\
\indent implementation,'' Ph.D. thesis (University of California, Berkeley, 
2008).

\noindent $^5$P.~Petric \emph{et al.,} Proc.~SPIE {\bf 7271}, 727107 (2009).

\noindent $^6$E.~Slot \emph{et al.,} Proc.~SPIE {\bf 6921}, 69211P (2008).

\noindent $^7$V.~Dai and A.~Zakhor, 
IEEE Trans. Image Process., {\bf 15}, 2522 (2006).

\noindent $^8$G.~Cramer, H.~Liu, and A.~Zakhor,
Proc.~SPIE {\bf 7637}, 76371L (2010).

\noindent $^9$A.~Carroll \emph{et al.,} Proc.~SPIE {\bf 9049}, 904917 (2014).

\noindent $^{10}$C.-C.~Wu, J.~Yang, W.-C.~Wang, and S.-J.~Lin,
Proc.~SPIE {\bf 9423}, 94231P (2015).

\noindent $^{11}$J.~Yang and S.~A.~Savari, 
J.~Micro/Nanolithogr., MEMS, MOEMS, {\bf 10}, 043007 (2011).

\noindent $^{12}$J.~Yang and S.~A.~Savari, 
Proc.~SPIE {\bf 7970}, 79701U (2011).

\noindent $^{13}$J.~Yang and S.~A.~Savari, 
IEEE Data Compr.~Conf., 109 (2010).

\noindent $^{14}$J.~Yang and S.~A.~Savari, 
{\em Recent Advances in Nanofabrication Techniques and Applications}, \\
\indent edited by B.~Cui, (InTech - Open Access Publisher, Rijeka, Croatia, 
2011), pp.~95-110.

\noindent $^{15}$C.-K.~Tang, M.-S.~Su and Y.-C.~Lu, 
IEEE Signal Process. Let., {\bf 20}, 645 (2013).

\noindent $^{16}$F.~Krecinic, S.-J.~Lin, and J.~J.~H.~Chen, 
Proc.~SPIE {\bf 7970}, 797010 (2011).

\noindent $^{17}$I.~H.~Witten, A.~Moffat, and T.~C.~Bell, 
{\em Managing Gigabytes: Compressing and Indexing Documents and} \\
\indent {\em Images}, 2nd ed. (Morgan Kaufmann, San Francisco, 1999).

\noindent $^{18}$T.~C.~Bell, I.~H.~Witten, and J.~Cleary, 
{\em Text Compression} (Prentice Hall, New Jersey, 1990).

\noindent $^{19}$J.~Ziv and A.~Lempel,
IEEE Trans. Inf. Theory {\bf IT-23}, 337 (1977).

\noindent $^{20}$J.~Ziv and A.~Lempel,
IEEE Trans. Inf. Theory {\bf IT-24}, 530 (1978).

\noindent $^{21}$K.~Sayood, {\em Introduction to Data Compression},
4th ed. (Morgan Kaufmann, Waltham, 2012).

\noindent $^{22}$D.~A.~Huffman, Proc. IRE {\bf 40}, 1098 (1952).

\noindent $^{23}$H.~Liu, V.~Dai, A.~Zakhor, and B.~Nikolic,
J.~Micro/Nanolithogr., MEMS, MOEMS, {\bf 6}, 013007 (2007).

\noindent $^{24}$H.~Liu, ``Architecture and hardware design of lossless 
compression algorithms for direct-write maskless \\
\indent lithography systems,'' Ph.D. thesis (University of California, 
Berkeley, 2008).

\noindent $^{25}$S.~W.~Golomb, IEEE Trans.~Inf.~Theory {\bf IT-12}, 399
(1966).

\noindent $^{26}$S.~M.~Rubin, \emph{Computer Aids for VLSI Design}, 
2nd ed. (Addison-Wesley, Boston, 1987).

\noindent $^{27}$A.~Moffat, R.~M.~Neal, and I.~H.~Witten, 
ACM Trans.~Inf.~Syst. {\bf 16}, 256 (1998). 

\noindent $^{28}$V.~Sze and A.~P.~Chandrakasan, 
IEEE J.~Solid-State Circuits {\bf 47}, 8 (2012).

\noindent $^{29}$J.~Zhou, D.~Zhou, J.~Zhu, and S.~Goto, 
IEEE T.~VLSI~Syst.~{\bf PP}, 1 (2015).

\noindent $^{30}$Y.-H. Chen and V.~Sze, 
IEEE T.~Circ.~Syst.~Vid.~{\bf 25}, 856 (2015).

\noindent $^{31}$J.~Yang and S.~A.~Savari, 
Proc.~SPIE {\bf 8352}, 83520K (2012).

\noindent $^{32}$J.~Yang and S.~A.~Savari, 
Proc.~SPIE {\bf 8680}, 86800Q (2013).

\noindent $^{33}$J.~Yang, S.~A.~Savari, and H.~R.~Harris,
J.~Micro/Nanolithogr., MEMS, MOEMS {\bf 12}, 033018 (2013).

\noindent $^{34}$P.~Elias, IEEE Trans.~Inf.~Theory {\bf IT-21}, 194 (1975).

\noindent $^{35}$T.-C.~Chen {\em et al.}, IEEE Circuits Devices Mag.
{\bf 22}, 22 (2006).

\noindent $^{36}$R.~Moussalli, W.~Najjar, X.~Luo, and A.~Khan,
Proc.~Int.~C.~Tools~Art {\bf 21}, 65 (2013).

\noindent $^{37}$E-mail correspondences from Chi-Cheng Ju, MediaTek Incorp.,
21 May 2014 and 25 June 2014.

\noindent $^{38}$J.~Fowers, J.-Y.~Kim, D.~Burger, and S.~Hauck,
Proc.~Int.~C.~Tools~Art {\bf 23} (2015).

\noindent $^{39}$A.~Martin, D.~Jamsek, and K.~Agarwal, ``FPGA-Based
Application Acceleration: Case Study with GZIP \\
\indent Compression/Decompression Streaming Engine,'' in Special Section 7C, 
IEEE IC CAD (2013).

\noindent $^{40}$H.~P.~Hofstee, ``The big deal about big data,'' Keynote 
presentation, {\em IEEE NAS Conf.}, Xi'An, China \\
\indent  17 July 2013, see \\
\indent http://www.nas-conference.org/NAS-2013/conference\%20speech/NAS\%20XiAn\%20Big\%20Data.pdf.

\noindent $^{41}$A.~W.~Paeth, {\em Graphic Gems II}, edited by J.~Arvo,
(Academic Press, San Diego, 1991), pp.~93-100.

\noindent $^{42}$A.~D.~Wyner and J.~Ziv, P.~IEEE {\bf 82}, 872 (1994).

\noindent $^{43}$B.~J.~Lin, Proc.~SPIE {\bf 9426}, 942602 (2015).

\subsubsection*{Tables}
\begin{table}
\caption{This table offers the compression ratios for the image 
compression block.}
\label{table:compression-denny}
\begin{center}
\begin{tabular}{r r r r r r r r}
\hline
\hline
Layer & PNG & BC4 & LDE & Corner2-EPC & Corner2-EPC & Corner2-EPC & P-EPC \tabularnewline
\mbox{} & \mbox{} & \mbox{} & \mbox{} & (plain) & (AC) & (deflate)
& (deflate) \tabularnewline
\hline
1 & 695.8 & 909.6 & 1,310.4 & 1,295.1 & 1,690.3 & 2,766.7 & 3,221.9
\tabularnewline
2 & 998.2 & 1339.7 & 9,832.7 & 44,661.2 & 64,367.8 & 395,852.5 & 500,239.1
\tabularnewline
3 & 364.2 & 436.4 & 320.8 & 237.1 & 327.9 & 793.5 & 835.7
\tabularnewline
4 & 395.9 & 503.0 & 357.2 & 250.3 & 347.5 & 852.7 & 874.4
\tabularnewline
5 & 272.6 & 308.1 & 283.0 & 249.9 & 335.1 & 488.2 & 549.7
\tabularnewline
6 & 952.7 & 1364.0 & 113,246.6 & 5,520,602.2 & 6,229,328.2 & 6,447,136.9 & 6,931,884.0
\tabularnewline
7 & 1026.6 & 1363.6 & 110,956.9 & 1,132,605.1 & 1,575,966.8 & 2,604,351.9 & 2,664,568.1
\tabularnewline
8 & 247.0 & 279.6 & 210.0 & 153.7 & 207.3 & 404.0 & 415.1
\tabularnewline
9 & 482.2 & 568.5 & 535.5 & 364.4 & 520.6  & 1,020.3 & 1,027.9
\tabularnewline
10 & 429.5 & 538.7 & 605.8 & 465.3 & 639.1 & 782.6 & 1,001.0
\tabularnewline
11 & 1026.6 & 1363.9 & 80,659.7 & 2,044,214.1 & 2,908,329.9 & 7,146,826.1 & 7,495,451.8
\tabularnewline
12 & 437.3 & 498.5 & 418.6 & 294.0 & 407.4 & 838.0 & 869.5
\tabularnewline
13 & 836.0 & 1105.7 & 1,864.4 & 2,663.5 & 3,739.3 & 6,273.0 & 6,341.9
\tabularnewline
14 & 766.2 & 1114.3 & 1,469.3 & 3,284.2 & 4,551.7 & 6,694.0 & 6,918.9
\tabularnewline
15 & 1026.6 & 1364.1 & 117,504.5 & 6,065,398.5 & 6,778,974.8 & 6,680,728.8 & 7,435,004.6
\tabularnewline
16 & 770.8 & 1019.1 & 1,420.4 & 1,935.7 & 2,742.8 & 5,089.4 & 5,114.9
\tabularnewline
17 & 1025.4 & 1362.9 & 83,980.7 & 717,463.5 & 911,008.5 & 1,403,258.1 & 1,501,531.9
\tabularnewline
18 & 1025.4 & 1363.4 & 85,961.8 & 1,074,522.8 & 1,268,143.8 & 1,477,468.9 & 1,790,175.9
\tabularnewline
19 & 1026.6 & 1364.3 & 116,040.3 & 14,870,009.2 & 15,895,527.1 & 18,438,811.4 & 20,042,186.4
\tabularnewline
20 & 1025.4 & 1362.9 & 83,866.1 & 704,309.1 & 896,829.4 & 1,390,558.9 & 1,487,000.9
\tabularnewline
21 & 1026.6 & 1364.3 & 118,258.2 & 11,381,982.4 & 12,458,656.4 & 11,670,133.8 & 13,170,579.6
\tabularnewline
22 & 1016.2 & 1364.3 & 119,592.8 & 17,072,973.6 &  17,729,626.4 & 16,174,396.0 & 17,729,626.4
\tabularnewline
23 & 1026.6 & 1364.3 & 118,258.2 & 11,381,982.4 & 12,458,656.4 & 11,670,133.8 & 13,170,579.6
\tabularnewline
24 & 926.3 & 1331.3 & 20,586.4 & 26,130.6 & 38,607.2 & 73,397.1 & 75,879.9
\tabularnewline
25 & 946.7 & 1362.7 & 88,682.2 & 863,240.2 & 1,322,726.8 & 1,726,480.5 & 1,866,276.5
\tabularnewline
All & 643.4 & 809.0 & 1,083.5 & 860.6 & 1,181.2 & 2,213.5 & 2,377.1
\tabularnewline
\hline \hline
\end{tabular}
\par\end{center}
\end{table}

\bigskip

\begin{table}
\caption{This table offers some statistics about the encoding and decoding 
times for the image compression block.}
\begin{center}
\begin{tabular}{ r r r r r r r }
\hline \hline
\multirow{2}{*}{Algorithm}
&\multicolumn{3}{c}{Encoding Time (seconds)}
&\multicolumn{3}{c}{Decoding Time (seconds)}\\
\cline{2-7}
 & Best & Worst & Average & Best & Worst & Average 
\tabularnewline
\hline
PNG & 14.24 & 16.62 & 15.69 & 0.83 & 3.75 & 1.60 
\tabularnewline
BC4 & 3464.49 & 3581.30 & 3500.66 & 72.39 & 113.67 & 89.32 
\tabularnewline
LDE & 1.75 & 2.29 & 1.87 & 0.88 & 4.99 & 2.014 
\tabularnewline
Corner2-EPC (Plain) & 3.75 & 4.12 & 3.84 & 2.00 & 2.69 & 2.19 
\tabularnewline
Corner2-EPC (AC) & 3.75 & 5.04 & 4.04 & 1.99 & 3.69 & 2.41 
\tabularnewline
Corner2-EPC (Deflate) & 3.26 & 3.74 & 3.37 & 1.97 & 2.63 & 2.19 
\tabularnewline
P-EPC (Deflate) & 7.25 & 7.91 & 7.42 & 5.13 & 6.00 & 5.50 
\tabularnewline
\hline \hline
\end{tabular}
\end{center}
\end{table}

\bigskip

\begin{table}
\caption{This table offers the compression ratios for the BFSK 
transmitter circuit.}
\label{table:compression-chipframe}
\begin{center}
\begin{tabular}{r r r r r r r r}
\hline
\hline
Layer & PNG & BC4 & LDE & Corner2-EPC & Corner2-EPC & Corner2-EPC & P-EPC \tabularnewline
\mbox{} & \mbox{} & \mbox{} & \mbox{} & (plain) & (AC) & (deflate)
& (deflate) \tabularnewline
\hline
1 & 1014.4 & 1357.3 & 54948.4 & 160735.2 & 243394.2 & 347199.8 & 415734.3
\tabularnewline
2 & 958.6 & 1320.3 & 5116.2 & 25191.7 & 36472.7 & 49379.3 & 60172.4
\tabularnewline
3 & 899.1 & 1230.4 & 2102.1 & 7327.9 & 10033.9 & 16600.7 & 20200.2
\tabularnewline
4 & 948.0 & 1307.3 & 3482.1 & 18693.6 & 25698.2 & 41625.2 & 50916.3
\tabularnewline
5 & 887.0 & 1227.7 & 1745.4 & 5501.9 & 7950.7 & 17830.3 & 21953.6
\tabularnewline
6 & 870.2 & 1205.8 & 1635.4 & 4800.1 & 6448.9 & 15171.1 & 18603.2
\tabularnewline
7 & 412.6 & 613.5 & 77.6 & 54.5 & 117.7 & 1483.7 & 1889.5
\tabularnewline
8 & 761.6 & 1035.4 & 871.6 & 2392.9 & 3251.0 & 5253.5 & 6428.3
\tabularnewline
9 & 712.9 & 1018.5 & 462.2 & 454.0 & 810.1 & 7799.9 & 8177.1
\tabularnewline
10 & 788.4 & 1083.4 & 780.6 & 2655.2 & 3652.9 & 7294.8 & 8397.7
\tabularnewline
11 & 991.1 & 1349.3 & 28198.8 & 76023.5 & 131744.4 & 390702.9 & 452622.2
\tabularnewline
12 & 751.3 & 1063.4 & 607.8 & 573.2 & 1020.4 & 10684.0 & 11297.3
\tabularnewline
13 & 885.3 & 1230.5 & 1628.8 & 5741.0 & 8055.5 & 20852 & 24221.6
\tabularnewline
14 & 753.6 & 1064.2 & 635.4 & 595.8 & 1057.7 & 11648.5 & 12247.6
\tabularnewline
15 & 888.4 & 1234.5 & 1757.3 & 5987.9 & 8463.0 & 22302.9 & 25905.7
\tabularnewline
16 & 760.8 & 1082.9 & 673.8 & 653.0 & 1187.6 & 14305.8 & 14921.1
\tabularnewline
17 & 890.1 & 1239.8 & 1978.6 & 5890.1 & 8707.9 & 25277.4 & 29534.5
\tabularnewline
18 & 979.4 & 1319.8 & 5712.8 & 24011.9 & 34993.9 & 71174.4 & 83414.3
\tabularnewline
All & 808.6 & 1128.7 & 685.8 & 654.7 & 1293.1 & 10242.8 & 12172.3
\tabularnewline
\hline \hline
\end{tabular}
\par\end{center}
\end{table}

\bigskip

\begin{table}
\caption{This table offers some statistics on the encoding and decoding 
times for the BFSK transmitter circuit.}
\begin{center}
\begin{tabular}{ r r r r r r r }
\hline \hline
\multirow{2}{*}{Algorithm}
&\multicolumn{3}{c}{Encoding Time (seconds)}
&\multicolumn{3}{c}{Decoding Time (seconds)}\\
\cline{2-7}
 & Best & Worst & Average & Best & Worst & Average 
\tabularnewline
\hline
PNG & 98.34 & 109.51 & 102.73 & 6.64 & 27.41 & 16.70
\tabularnewline
BC4 & 23126.36 & 23751.60 & 23400.18 & 487.28 & 937.71 & 599.39
\tabularnewline
LDE & 11.79 & 19.62 & 12.66 & 6.77 & 43.14 & 19.05
\tabularnewline
Corner2-EPC(Plain) & 27.44 & 34.05 & 28.03 & 14.10 & 20.10 & 15.66
\tabularnewline
Corner2-EPC(AC) & 27.26 & 46.41 & 29.15 & 13.86 & 37.73 & 17.04
\tabularnewline
Corner2-EPC(Deflate) & 27.25 & 31.96 & 27.92 & 13.89 & 19.76 & 15.46
\tabularnewline
P-EPC(Deflate) & 51.16 & 55.67 & 52.33 & 36.62 & 41.95 & 38.20
\tabularnewline
\hline \hline
\end{tabular}
\end{center}
\end{table}
\newpage 

$\vspace*{6in}$

\newpage 

$\vspace*{4in}$

\subsubsection*{Figure Captions}
\noindent Fig.~1 (Color Online) 
The figure depicts data delivery for electron-beam
direct-write lithography systems. Reproduced with permission from 
J.~Vac.~Sci.~Technol., B~{\bf 32}, 06F502 (2014). Copyright 2014 American
Vacuum Society.

\noindent Fig.~2 (Color Online)
The figure illustrates the two-step transformation 
of an image. (a) The original image has pixel values in 
the range $\{0, \dots , 31\}$. (b) After the horizontal coding step, the 
intermediate image has pixel values in the range $\{-31, \dots , 31\}$. 
(c) After the vertical coding step, the transformed image has pixel values 
in the range $\{-62, ..., 62\}$.

\end{document}